\documentclass[aps,prl,twocolumn,groupedaddress]{revtex4-1}
\usepackage{graphicx}% Include figure files
\usepackage{dcolumn}% Align table columns on decimal point
\usepackage{bm}% bold math
\begin{document}

%\preprint{}

\title{Displacement induced electric force and natural self-oscillation of a free electron}


%\thanks{}
\author{Zhixian Yu}
\affiliation{Department of Physics and Astronomy, University of New Mexico, Albuquerque NM 87131 USA}
\affiliation{College of Physics Science, Qingdao University, Qingdao 266071 China}
\author{Liang Yu}
\email{lyuqdcn@gmail.com}
\affiliation{Laboratory of Physics, Jiaming Energy Research, Qingdao 266003 China}

\date{\today}

\begin{abstract}
We show that a kind of displacement induced temporary electric force of a single point charge can be derived by using Maxwell stress analysis.
This force comes from the variation of the charge's electric intensities that follow Coulomb's inverse square law, and it is a kind
of displacement dependent temporary restoring force. We also show the possible existence of natural self-oscillation of a free electron which is driven by this restoring self-force of its own electric fields.
\begin{description}
\item[PACS numbers]
03.50.De
\end{description}
\end{abstract}

%\keywords{}

\maketitle
\section{Introduction}
The distance dependence of the electrostatic force of Coulomb's
inverse square law has been tested in high precision through
different methods, see for example the discussion of Jackson\cite{Jackson3}
and experiment of Williams et al\cite{Hill}.

In this paper we want to show that some temporary displacement
dependent electric force may be generated in relation to the
Coulomb's law. This is a kind of restoring electric self-force. It is induced
by any displacement of the electric charge which changes the
distribution of its electric fields that follow Coulomb's law.
\section{The variation of distribution and magnitude of electric intensity of a charge subject to temporary displacement}
For a particle with electric charge $e$ at rest at origin $O$ first,
according to Coulomb's law the distribution of its spherically
symmetric electric intensity $E$ follows the inverse square law, that is
\begin{equation}
{\bf E}={e\over r_e^3}{\bf {r}_e},
\end{equation}
where $r_e$ is the magnitude of position vector from the particle to a
field point. Suppose this charge is shifted along $z$ axis by a displacement
$Z_e$ in a time interval $\triangle t$ and become stationary again
at a position $O_e$. Thus the electromagnetic fields depending on
its velocity and acceleration may be neglected first. By this
displacement $Z_e$, the electric intensity $E$ of Eq. (1) will change
in both magnitude and distribution. During this time interval any
variation signal of the electromagnetic fields will propagate first
from the origin $O$ to a spherical shell with radius $c\Delta{t}$
as shown in Fig. 1, where $c$ is velocity of light. At a position
$P$ on this spherical shell, the distance to origin $O$ is the
radius $r=c\Delta{t}$ and to the new origin $O_e$ is $[r_e]$. The
relation between $[r_e]$ and $r=c\Delta t$ is that
\begin{equation}
[r_e]^2=(c\Delta{t})^2+Z_e^2-2c\Delta{t}{Z_e}\cos \theta,
\end{equation}
where $\theta$ is the angle between $r$ and $z$ axis. Now the
electric field outside this spherical shell does not change, the magnitude of the electric intensity on the outer boundary of
this surface at $P$ is
\begin{equation}
E_{out}={e\over r^2}={e\over (c\Delta t)^2}.
\end{equation}
The magnitude of the electric intensity on the inner boundary of
this surface at $P$ is
\begin{equation}
E_{in}={e\over [r_e]^2}.
\end{equation}
Since by Eq. (2) $[r_e]^2\neq(c\Delta t)^2$, these two electric
intensities of Eq. (3) and Eq. (4) are not equal. But according to the
requirement of div${\bf E}=0$ in charge free space, they should be
equal in magnitude and direction. Although this requirement is
applied for the total electric field, it is also correct for the
Coulomb field alone here. Since $E_{out}$ of Eq. (3) is the primary
value, it must remain unchanged, thus $E_{in}$ should be modified to
fulfil the requirement of div${\bf E}=0$, that is to change $E_{in}$
to
\begin{figure}
\includegraphics[width=6cm]{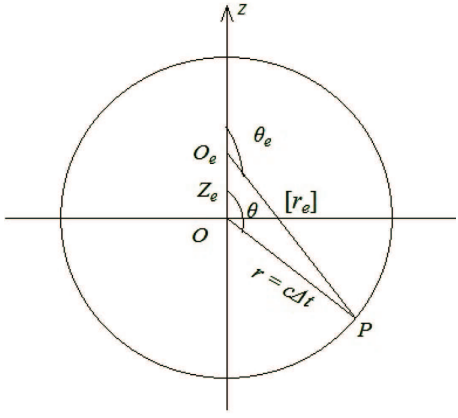}% Here is how to import EPS art
\caption{\label{fig:epsart} The relations between $r=c\Delta{t}$ and $[r_e]$.}
\end{figure}
\begin{equation}
E_{in}={e\over [r_e]^2}\cdot{[r_e]^2\over (c\Delta t)^2}={e\over
{r_e}^2}MF={e\over (c\Delta t)^2},
\end{equation}
where $MF$ is a kind of modification function. From Eq. (5) it is
defined as
\begin{equation}
MF={[r_e]^2\over (c\Delta t)^2}=1+{Z_e^2\over (c\Delta
t)^2}-{2Z_e\cos\theta\over c\Delta t}.
\end{equation}
This kind of modification may also be seen from the relation of the
solid angles $d\Omega$ and $d\Omega_e$ subtending to an area
element $ds$ at position $P$ from the origins at $O$ and $O_e$
respectively as shown in Fig. 2. For the two solid angles we have
\begin{equation}
d\Omega={ds\over r^2}={ds\over (c\Delta t)^2},
\end{equation}
\begin{equation}
d\Omega_e={ds\over [r_e]^2},
\end{equation}
and following Eq. (6) we have
\begin{equation}
{d\Omega\over d\Omega_e}={[r_e]^2\over r^2}={[r_e]^2\over (c\Delta
t)^2}=MF.
\end{equation}
Since the number of electric lines within solid angle $d\Omega_e$ should be the same as that within $d\Omega$, the area density of electric lines which represents the electric intensity will change by the ratio $MF$. This modification now not only modifies the electric intensity at position $P$, but should also be down to the whole solid angle $d{\Omega_e}$. Thus we may write the electric intensity within the solid angle $d{\Omega_e}$ as
\begin{equation}
E_e={e\over r_e^2}MF.
\end{equation}
Then when the charge is subject to a temporary displacement
$Z_e$ during time interval $\Delta t$, the electric intensities should be changed totally
according to Eq. (10) within the spherical surface of radius
$r=c\Delta t$.
\begin{figure}
\includegraphics[width=6cm]{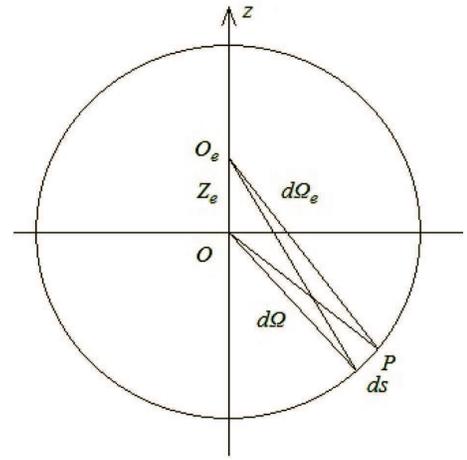}% Here is how to import EPS art
\caption{\label{fig:epsart} The relation between solid angles $d\Omega_e$ and $d\Omega$.}
\end{figure}
\section{Maxwell stress analysis of the displacement dependent electric force on a charge subject to a temporary displacement}
It is known that electromagnetic force on a system may be calculated
from the Maxwell stress on the surface that encloses this system. We
see from above that for a particle with charge $e$ and subject to a
temporary displacement $Z_e$, the distribution and magnitude of its
electric intensity follows the modified Coulomb's law as Eq. (10). Now
we want to show that this particle will get a displacement dependent
electric force generated by its own electric field, and this can be
calculated from the Maxwell stress surrounding this particle. Now we
may take a spherical surface with center at $O_e$ and radius $a$ surrounding this
particle, and according to Eq. (10), the electric intensity at this
surface $E_a$ as shown in Fig. 3 is

\begin{equation}
E_a={e\over a^2}MF.
\end{equation}

According to the theory of Maxwell stress, the electric force on
this particle may be calculated from the stress on the
above-mentioned surface. For any electric intensity in space there
is a corresponding Maxwell stress tensor $\bf T$ as\cite{Jackson1}
\begin{equation}
{\bf T}={1\over 4\pi}[{\bf EE}-{1\over 2}{\bf I}(E^2)],
\end{equation}
where $\bf I$ is the unit second rank tensor, the elements of $\bf
T$ are
\begin{equation}
T_{ij}={1\over 4\pi}[E_iE_j-{1\over 2}I_{ij}E^2],
\end{equation}
its force element $dF_i$ on an area element $ds$ at the surface is
\begin{equation}
dF_i=\sum_{j=1}^{3}T_{ij}ds_j,
\end{equation}
where $ds_j$ is the $j$ component of $ds$. The force $F_i$ on that surface is
\begin{equation}
F_i=\int dF_i.
\end{equation}
On the spherical surface surrounding the charged particle, the
distribution of electric intensity is Eq. (11) with directions along
the normal of the surface element $ds$. Using Eqs. (13), (14) and (15), the
total force along $z$ axis on the closed spherical surface is
\begin{eqnarray}
F_z&=&\oint(T_{zz}ds_z+T_{zx}ds_x+T_{zy}ds_y)\nonumber\\&=&\oint{1\over
4\pi}{1\over 2}(E_{z}^2-E_{x}^2-E_{y}^2)ds_z,
\end{eqnarray}
since the last two terms integrate to zero owing to their symmetry.
Here $ds_z=a^2\sin{\theta_e}\cos{\theta_e}d\theta_ed\phi$ and we may take $\theta\approx\theta_e$ when $Z_e$ is small. Substituting $MF$ of Eq. (6) into Eq. (16) and using $E_z=E_a\cos{\theta_e}$, $E_x=E_a\sin{\theta_e}\cos{\phi}$ and $E_y=E_a\sin{\theta_e}\sin{\phi}$, we have
\begin{widetext}
\begin{eqnarray}
F_z&=&\displaystyle\oint{1\over 4\pi}{1\over 2}{e^2\over
a^4}[1+{Z_e^2\over (c\Delta t)^2}-2{Z_e\cos\theta_e\over
{c\Delta t}}]^2(\cos^2\theta_e-\sin^2\theta_e\cos^2\phi-\sin^2\theta_e\sin^2\phi)a^2\sin\theta_e\cos\theta_ed\theta_ed\phi\nonumber\\
&=&\displaystyle\oint{e^2\over 4a^2}[1+{Z_e^4\over (c\Delta
t)^4}+{4Z_e^2\cos^2\theta_e+2Z_e^2\over ({c\Delta
t})^2}-{4Z_e\cos\theta_e\over c\Delta t}-{4Z_e^3\cos\theta_e \over
({c\Delta
t})^3}](2\cos^2{\theta_e}-1)\cos{\theta_e}\sin{\theta_e}d\theta_e.
\end{eqnarray}
\end{widetext}
Here we integrate $\phi$ from 0 to $2\pi$. The odd terms of
$\cos{\theta_e}$ integrate to zero. Since $Z_e$ is small, the important
term of above integration is the first order term of $Z_e$, those
higher order terms can be neglected, thus we have
\begin{equation}
F_z=\oint{e^2\over 4a^2}(-{4Z_e\cos\theta_e\over
c{\Delta}t})(2\cos^2\theta_e-1)\cos\theta_e\sin\theta_ed\theta_e,
\end{equation}
\begin{figure}
\includegraphics[width=8cm]{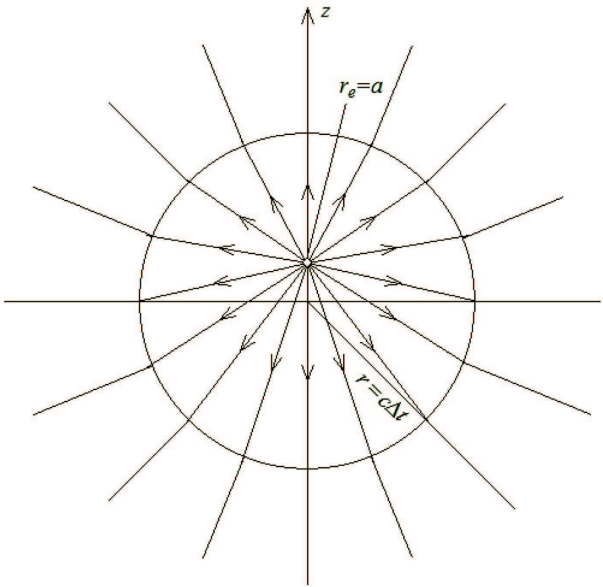}% Here is how to import EPS art
\caption{\label{fig:epsart} The distribution of electric intensities and electric field lines between two spherical surface with radii $r=c\Delta{t}$ and $r_e=a$.}
\end{figure}
thus
\begin{equation}
F_z=-{2e^2\over 15{a^2}}{Z_e\over c\Delta t}.
\end{equation}
This displacement dependent $F_z$ is independent of the positive or
negative sign of charge $e$ since it's proportional to $e^2$. This
is a kind of displacement dependent restore force induced by the
variations of the charge's own electric intensity which follows
Coulomb's inverse square law. The magnitude of this force is
dependent on the radius $a$ of the spherical surface boundary of the
charge. Here this force is also a temporary force. After the time interval $\Delta t$, as time goes on, $\Delta t$ increases to ${\Delta t}+t$,
where $t$ is the additional time, then Eq. (19) will change to
\begin{equation}
F_z=-{2e^2\over 15{a^2}}{Z_e\over c({\Delta t}+t)}.
\end{equation}
When $t$ becomes large, $F_z$ reduces to zero.

This problem cannot be treated by usual methods such as the time
dependent Coulomb's law\cite{Griffiths}, the Lienard-Wiechart potential\cite{Panofsky}
or the Feynman's formula\cite{Feynman} for a charge undergoing an arbitrary
translation motion, since the variations of internal structure of
the electric fields are not taken into account in these methods.
Although the retarded time calculations are used in these methods,
the historical conditions of the electric fields are ignored. The
problem here is a kind of hereditary electromagnetism\cite{Volterra}. We do
not use the integro-differential equation of functional analysis for
this problem but the transit effect of displacement is treated in
similar way\cite{Volterra}. Similar analysis about the radiation of electric charge induced by its acceleration via the variation of electric field lines was given by others\cite{Ohanian}. Our discussions here emphasize only on the effect of displacement. The effect of velocity dependent magnetic fields is neglected and the equation of static cases is used as Eq. (14).
\section{Possible effect of the displacement dependent restore force on the dynamics of a free electron}
The displacement dependent restoring force of Eq. (19) is a kind of
self-induced force. For a single free electron with charge $e$ and
mass $m_e$ which is not subject to any external force, this
self-induced force will affect its dynamical motion. We may take
$M_0r_0$ as the radius $a$ above, where $M_0$ is an undetermined
numerical constant, $r_0$ is the classical radius of electron which
is defined as\cite{Feynman}
\begin{equation}
r_0={e^2\over m_ec^2}\approx0.82\times10^{-13}cm.
\end{equation}
Substituting $M_0r_0$ into Eq. (19), we get
\begin{equation}
F_z=-{2e^2\over 15{(M_0r_0)^2}}{Z_e\over c\Delta t}.
\end{equation}
For an electron possessing self-sustained harmonic oscillation with
angular frequency $\omega$, the time interval $\Delta t$ may be taken as $\pi/\omega$,
which is half of its period. Then we have
\begin{equation}
F_z=-{2e^2\over 15{(M_0r_0)^2}}{\omega Z_e\over
c\pi}=-{2m_ec\omega{Z_e}\over 15{M_0^2}\pi{r_0}},
\end{equation}
since $m_e={e^2/(r_0c^2)}$ according to Eq. (21). If we take ${2c/
(15M_0^2\pi{r_0})}=\omega$, then
\begin{equation}
F_z=-m_e\omega^2Z_e,
\end{equation}
which is a standard equation of harmonic oscillation of the free
electron. This displacement dependent restore force implies that the
electron may have a kind of natural self-oscillation as the ``mono-electron
oscillator" in a Penning trap of the experiment of Dehmelt et
al\cite{Wineland}.

The oscillating electron will have oscillating static fields,
velocity dependent and acceleration dependent fields, which may
store energy and exchange energy between each other\cite{Marion,Mandel}. They
need not always radiate out their energy through acceleration
dependent electromagnetic fields if these fields have standing wave
mode solutions\cite{Bateman,Adler}.

\end{document}